\documentclass{webofc}

\usepackage[varg]{txfonts}   
\usepackage{hyperref}
\usepackage{url}
\hypersetup{colorlinks=true,citecolor=blue,urlcolor=blue,linkcolor=blue}
\begin{document}
\title{Recent theoretical developments in quarkonia production in relativistic heavy ion collisions}
%

\author{\firstname{Jiaxing} \lastname{Zhao}\inst{1,2,3}\fnsep\thanks{\email{jzhao@itp.uni-frankfurt.de}}
}
\institute{
Helmholtz Research Academy Hesse for FAIR (HFHF), GSI Helmholtz Center for Heavy Ion Physics, Campus Frankfurt, 60438 Frankfurt, Germany  
\and
Institut f\"ur Theoretische Physik, Johann Wolfgang Goethe-Universit\"at,Max-von-Laue-Straße 1, D-60438 Frankfurt am Main, Germany
\and
SUBATECH, Nantes University, IMT Atlantique, IN2P3/CNRS
4 rue Alfred Kastler, 44307 Nantes cedex 3, France
}

\abstract{
Quarkonia are golden probes of heavy ion collisions that have attracted much attention from both experimental and theoretical perspectives. This paper will review recent theoretical studies on quarkonium thermal properties, with a particular focus on the heavy quark finite-temperature potential obtained by Lattice QCD and other approaches.  Moreover, it will examine the advancements in the real-time evolution of quarkonia in heavy ion collisions.
}
\maketitle

\section{Introduction}
\label{sec.intro}

Quarkonia, the bound state of a heavy quark (HQ) and its antiquark, are commonly referred to as charmonium and bottomonium, respectively. Due to their large mass, they are mostly produced in the early stages of heavy ion collisions and thus undergo all stages of evolution. Relativistic heavy ion collisions produce a hot and dense deconfined matter, known as the quark-gluon plasma (QGP). Accordingly, the production of quarkonium in relativistic heavy ion collisions depends on its thermal properties in the hot QGP medium and its real-time dynamical evolution framework. From a theoretical perspective, HQs with large mass and scale separation facilitate the advancement of the heavy quark effective theory. Furthermore, the QGP offers an optimal thermal environment for investigating quantum chromodynamics (QCD) at finite temperature, encompassing both perturbative and non-perturbative aspects. 

\section{Quarkonium thermal properties}
\label{sec.pro}

The hierarchy of the heavy quark energy scales, namely $m_Q\gg m_Qv\gg m_Qv^2$, in vacuum allows for the quarkonia to be described by a series of effective theories based on QCD, such as, the non-relativistic QCD (NRQCD) and potential non-relativistic QCD (pNRQCD)~\cite{Brambilla:1999xf}. The equation of motion in pNRQCD returns to a Schr\"odinger-like equation with the transition between the color singlet state and the octet state~\cite{Brambilla:1999xf}, where the leading order potential for the singlet state is $V(r)=-C_F\alpha_s/r$ with $C_F=(N_c^2-1)/(2N_c)$ and the number of colors $N_c=3$. Moreover, phenomenological studies and non-perturbative lattice simulations indicate the presence of a linear confinement potential between a heavy quark-antiquark pair in the long range. The full range potential can be described by a Cornell form, namely, $V(r)=-\alpha/r+\sigma r +\rm constant$. The two-body Schr\"odinger equation with the Cornell potential can provide a successful description of the quarkonium mass spectra, see the review pape~\cite{Zhao:2020jqu}. 

\begin{figure}[!htb]
\centering
\includegraphics[width=6cm]{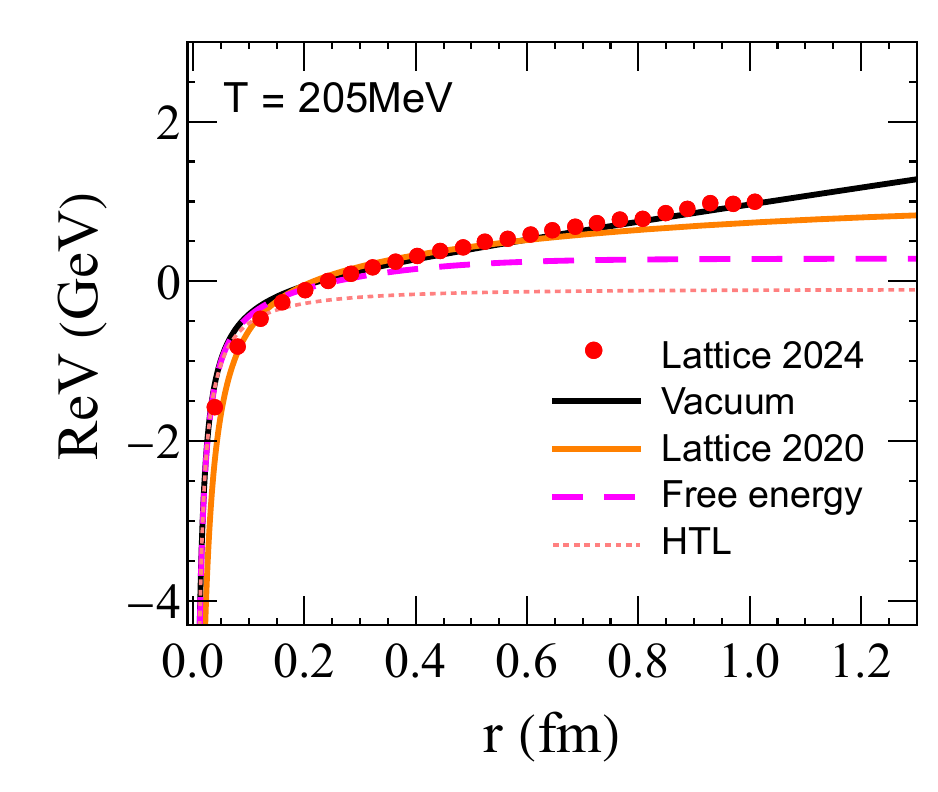}\includegraphics[width=6cm]{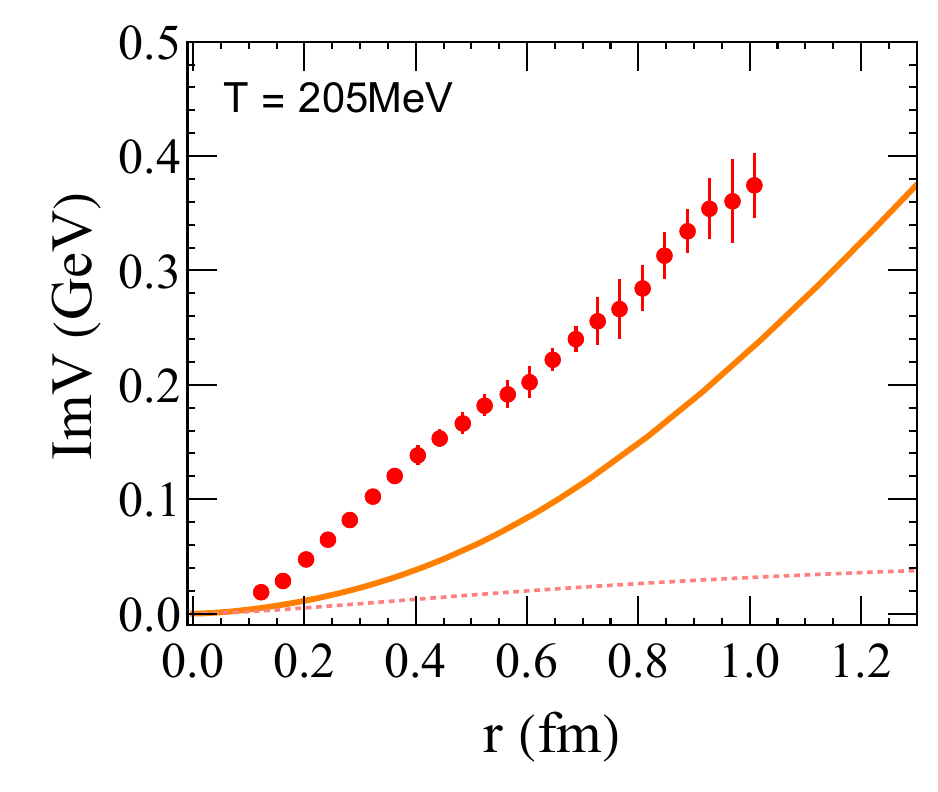}
\caption{Comparison of the different real (left panel) and imaginary (right panel) part potentials at T=205 MeV. The results of lattice 2020 is from Ref.~\cite{Lafferty:2019jpr} and lattice 2024 is from Ref.~\cite{Bazavov:2023dci}.}
\label{fig.1} 
\end{figure}

To study the evolution and production of quarkonium in relativistic heavy ion collisions, it is essential to calculate its thermal properties in the hot QCD medium. All in-medium properties of the quarkonium are encoded in its spectral function. The quarkonium bound state manifests as a peak in the spectral function. As the temperature increases, the peak shifts in position and broadens. From the perturbative perspective, the mass shift can be attributed to the static color screening effect in the hot medium, whereas the width is a consequence of the thermal decay of the color singlet octet and the Landau damping, which correspond to gluon dissociation and inelastic scattering, respectively. Meanwhile, the in-medium properties can be incorporated into a temperature-dependent HQ potential. If the HQs interact with the medium for a very long time, the potential approaches the free energy~\cite{Kaczmarek:2004gv}. But in heavy ion collisions, the focus is on the HQ potential over a timescale that is comparable to the internal time scale of quarkonium. In the weak coupling regime, the potential is given by the hard thermal loop (HTL)~\cite{Laine:2006ns}, which has a complex form. In the strong coupling regime, lattice QCD has been employed to simulate the Euclidean Wilson loop, which can be related to the spectral function via a Laplace transform.  Nevertheless, the reconstruction of spectral functions from Euclidean correlation functions is an ill-posed inverse problem. A significant discrepancy is attributable to the extraction strategy, as shown in the recent 2+1 flavor lattice QCD study by the HotQCD Collaboration~\cite{Bala:2021fkm}! Assuming a Lorentzian form for the spectral functions, the extracted HQ potential exhibits no screening but displays a substantial imaginary component~\cite{Bazavov:2023dci}. The real and imaginary parts of the potentials at T=205 MeV from the aforementioned approaches are clearly illustrated in Fig.~\ref{fig.1}. 

This no-screening HQ potential is very different from the free energy and previous lattice results. The no-screening or less-screening potential has been demonstrated in many prior studies. For instance, it has been extracted from bottomonium mass and width~\cite{Shi:2021qri}, from bottomonium experimental observables~\cite{Du:2019tjf}, and by fitting the Wilson line correlators and EOS in the T-matrix approach~\cite{ZhanduoTang:2023pla}. Furthermore, the HQ potential has been studied with the HTL resumed perturbation method within the Gribov-Zwanziger approach, which enables the study of confinement at the perturbative level~\cite{Wu:2022nbv}. This ``new'' HQ potential suggests a markedly different picture of quarkonium melting in the QGP, with dynamic dissociation playing a dominant role rather than the color screening that has been recognized for many years~\cite{Matsui:1986dk}. The question of how to detect this potential in experiments and which observable it is sensitive to remains unanswered. What is needed is a real-time study of the evolution of quarkonium in the QGP.

\section{Real-time evolution in hot QCD medium}
\label{sec.evo}

The question can be posed as to whether a quarkonium can be defined as a classical particle or a quantum object. If a quarkonium is a very tightly bound state with a narrow width, it can be treated as a classical point particle. Moreover, if one assumes that the quarkonium is entirely dissociated in the QGP and is produced exclusively at the QCD phase boundary, The statistical hadronization model provides a description of the yield of quarkonium. A core-corona approach has recently been employed to study the transverse momentum spectra and the collective flow of quarkonium~\cite{Andronic:2023tui}. This framework provides a comprehensive description of the spectra and collective flow of quarkonium, particularly in the low $p_T$ region~\cite{Andronic:2023tui}.
Nevertheless, the tightly bound quarkonium with a high dissociation temperature can survive in the QGP medium. Therefore, the real dynamics of the quarkonium is only partially dissociated and accompanied by the regeneration within the QGP. In this case, the classical transport equation is employed to describe the quarkonium real-time evolution, such as the THU model (Boltzmann equation)~\cite{Yan:2006ve} and the TAMU model (rate equation)~\cite{Zhao:2011cv}, which have successful explained the quarkonium observables from RHIC to LHC energies. Recently, the non-thermal charm distribution and the space-momentum correlation have been considered in the context of regeneration, which appears to significantly enhance the $v_2$ of $J/\psi$ in the intermediate $p_T$ region. These classical transport models have been extended to study the directed flow $v_1$ and the triangle flow $v_3$, the quarkonium polarization, and also the quarkonium-like states such as $B_c$ and $X(3872)$.

It is therefore pertinent to inquire as to the significance of quantum effects in quarkonium transport, and to elucidate the nature of these effects. Two principal quantum effects have been identified. One such effect is that of quantum coherence and decoherence. In the classical transport model, it is typically assumed that the quarkonium is produced before it enters the hot medium. In the quantum view, however, the initially produced state is a compact wave packet that can be treated as a superposition of different quarkonium eigenstates. It takes time for this compact wave packet to decohere into quarkonium eigenstates, which is called ``formation time'' and is sometimes included in a phenomenological parameter in the classical transport approach. Nevertheless, further investigation into the manner in which the system decoheres within a hot medium would be beneficial. Another quantum effect is the decay process, which results in a significant width compared to that of a classical particle

\begin{figure}[!htb]
\centering
\includegraphics[width=13cm]{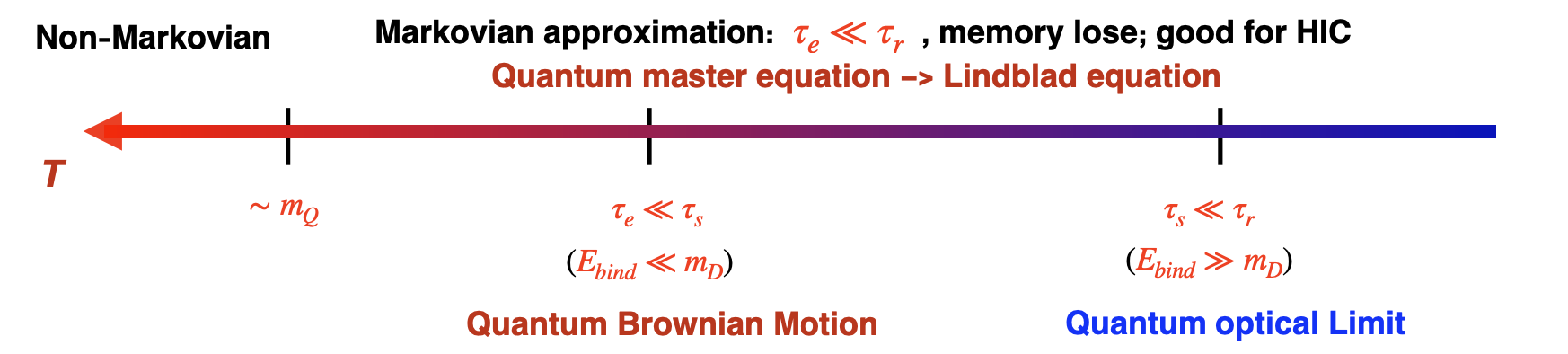}
\caption{Sketch of different time scales along the temperature line.}
\label{fig.2} 
\end{figure}

Assuming that the quarkonium is a quantum wavefunction, its real-time evolution can be described by the time-dependent Schr\"odinger equation with a temperature-dependent complex HQ potential. This framework has been employed to elucidate the evolution and production of bottomonium in heavy ion collisions~\cite{Islam:2020gdv,Chen:2024iil,Kumar:2023wvz}. Nevertheless, the wavefunction is employed solely for the pure state $\psi_i$. In order to address the mixed state (color octet and color singlet state), the density matrix is introduced. The total density operator for the system is given by $\rho_{tot}=\sum_ip_i|\psi_i><\psi_i|$. The evolution of the density operator is governed by the von Neumann equation, ${d\hat \rho_{tot}\over dt}=-i[\hat H_{tot}, \hat \rho_{tot}]$. A heavy quark pair $Q\bar Q$ in a hot QCD medium can be treated as an open quantum system (OQS). The total Hamiltonian $\hat H_{tot}$ can be separated into the environment part, the $Q\bar Q$ subsystem part, and the interaction part. The system exhibits three principal time scales:
\begin{itemize}
\item Environment relaxation time scale: $\tau_e\sim {1/ (\pi T)}$;
\item Intrinsic time scale of subsystem: $\tau_s\sim {1/ E_{\rm bind}}$;
\item Subsystem relaxation time scale: $\tau_r \sim {1/ \eta}\approx {m_Q/T^2}$.
\end{itemize}
We can see that these time scales are related to the temperature of the QGP. The very high temperature regime, where $T\gg m_Q$, is called the non-Markovian regime. In the relativistic heavy ion collisions, the typical temperature of the QGP medium is much smaller than the heavy quark mass, i.e. the relaxation time of the environment is much smaller than the relaxation time of the subsystem, $\tau_e\ll \tau_r$. In this instance, it is possible to trace the degrees of freedom of the environment and simplify the evolution equation of the density operator to the Lindblad form. This approximation is also referred to as the Markovian approximation as illustrated in Fig.~\ref{fig.2}. 

Furthermore, if the relaxation time of the environment is significantly shorter than the intrinsic time scale of the subsystem $\tau_e\ll \tau_s$, indicating that the binding energy $E_{\rm bind}$ of the quarkonium is less than the Debye screening mass $m_D$, the system is in the quantum Brownian motion region. This region is characterized by high temperatures, and the quarkonium exhibits a broad peak and a short quantum decoherence time. In the low-temperature region, the quarkonium exhibits a large binding energy and a narrow width. In this region, the intrinsic time of the subsystem is much smaller than the relaxation time of the subsystem, $\tau_s\ll \tau_r$, which is typically referred to as the Quantum Optical Limit. In recent years, significant theoretical advances have been made in these two regions, based on the first-principle QCD and OQS approachs, see recent review papers~\cite{Yao:2021lus,Akamatsu:2020ypb,Andronic:2024oxz}. I will show them separately below.
\begin{itemize}
\item \textit{Duke approach}. It works in the Quantum Optical Limit and based on the pNRQCD which means with the dipole approxiamtion. Following a semi-classical expansion, it was demonstrated that the Lindblad equation is equivalent to the classical Boltzamann equation~\cite{Yao:2018nmy}. This establishes a connection between the OQS and the classical transport equation. This approach has been employed to elucidate the production of bottomonium in heavy ion collisions.

\item \textit{Munich-Kent approach}. It is based on the pNRQCD as well but works in the Quantum Brownian Motion regime with the scale hierarchy, $m_Q\geq 1/a_0 \gg \pi T\sim m_D \gg E_{\rm bind}$. The Lindblad operators can be obtained by expanding the self-energy and transition function at the leading order or the next-to-leading order of $E_{\rm bind}/T$~\cite{Brambilla:2021wkt,Brambilla:2022ynh}. In this instance, the Lindblad operators are associated with two coefficients:
$\kappa$, which is the heavy quark momentum diffusion coefficient, and $\gamma$, which is the dispersive quantity. $\kappa$ and $\gamma$ are respectively related to the thermal width and mass shift of the bottomonium.
Recently, the researchers have implemented the quantum jumps and utilized the novel framework to investigate the bottomonium production in heavy ion collisions. The results demonstrated that the bottomonium $R_{AA}$ and the double ratio can be well described with quantum jumps and without color screening~\cite{Brambilla:2023hkw}. This result may support the aforementioned ``new'' potential. 

\item \textit{Nantes approach}. It works in Quantum Brownian Motion regime but based on NRQCD with the scale, $m_Q\gg T\sim m_D\geq E_{\rm bind}$, that means beyond the dipole approxiamtion. The Lindblad operators $\mathcal{L}_i$ are given by the expansion of the $\tau_e/\tau_s$~\cite{Blaizot:2017ypk,Delorme:2022hoo}. A numerical solution of the quantum master equations in one dimension has recently been carried out using different initial states and medium configurations. Furthermore, the contributions of the various operators that govern the evolution are also examined~\cite{Delorme:2024rdo}.

\item \textit{Osaka approach}. It works in Quantum Brownian Motion regime as well and based on NRQCD. In the weak coupling, the Lindblad equation is governed by two quantities $V(r)$ and $D(r)$, which are defined in terms of gluon two-point functions. In the HTL approximation, the explicit forms of $V(r)$ and $D(r)$ correspond to the real and imaginary part potentials, respectively~\cite{Akamatsu:2012vt}. Beyond the weak coupling, the assumption is made that the real and imaginary potentials can be derived from lattice QCD. The related study and the rationale behind the dipole approximation were recently investigated~\cite{Miura:2022arv}.
\end{itemize}

In the heavy ion collisions, the created QGP expands rapidly and the temperature drops gradually. The system transitions from the quantum Brownian motion regime at high temperatures to the quantum optical limit region at low temperatures. It is therefore necessary to construct a framework that connects these two regimes. Moreover, these studies concentrate on a single pair of heavy quarks, which is likely a reasonable approximation for the production of bottomonium.  
For charmonium, regeneration from many uncorrelated charm quarks plays an important role at top RHIC and LHC energies. Consider these two requirements,
A microscopic model based on the density matrix formalism has been developed to describe both charmonium~\cite{Villar:2022sbv,Song:2023ywt} and bottomonium~\cite{Song:2023zma} production in heavy ion collisions. In this framework, a quarkonium is represented as a Wigner density, which is a fully quantum object. The only approximation is to simulate the N-body Wigner density as a classical phase-space distribution. A unified framework for both open heavy and quarkonium production, based on EPOS4, will be available in the near future.

\section{Summary}
\label{sec.sum}
In conclusion, the recent advancements in the theoretical understanding of the thermal properties of quarkonium and its real-time evolution in the hot QCD medium are reviewed. 
The recent lattice QCD along with many other research efforts, indicate that the HQ in-medium potential exhibits either no color screening effect or a diminished one, accompanied by a pronounced imaginary component. This may challenge our current understanding of quarkonium dissociation in the QGP. 
The dynamical scattering plays a dominant role instead of the static color screening. 
The classical transport models can describe the experimental data very well and have been extended to many interesting studies in recent years. The creation of a comprehensive, first-principles-based framework for the real-time evolution of quarkonium is a crucial step in advancing our understanding of QCD. Furthermore, it can more effectively address the limitations of classical transport models. Additionally, there have been significant advancements in the field of quarkonium real-time evolution, particularly in the context of the NRQCD or pNRQCD formalism and based on the the open quantum system framework.

{\bf Acknowledgement}: The work is supported by the European Union's Horizon 2020 research and innovation program under grant agreement No 824093 (STRONG-2020), and the Deutsche Forschungsgemeinschaft (DFG, German Research Foundation) through the grant CRC-TR 211 ’Strong-interaction matter under extreme conditions’ - Project number 315477589 - TRR 211. 

%
%

\end{document}